\documentstyle[12pt,html,epsf,psfig]{article}

\begin{document}
\title{MATTERS OF GRAVITY, The newsletter of the APS Topical Group on 
Gravitation}
\begin{center}
{ \Large {\bf MATTERS OF GRAVITY}}\\ 
\bigskip
\hrule
\medskip
{The newsletter of the Topical Group on Gravitation of the American Physical 
Society}\\
\medskip
{\bf Number 23 \hfill Spring 2004}
\end{center}
\begin{flushleft}

\tableofcontents
\vfill
\section*{\noindent  Editor\hfill}

Jorge Pullin\\
\smallskip
Department of Physics and Astronomy\\
Louisiana State University\\
Baton Rouge, LA 70803-4001\\
Phone/Fax: (225)578-0464\\
Internet: 
\htmladdnormallink{\protect {\tt{pullin@phys.lsu.edu}}}
{mailto:pullin@phys.lsu.edu}\\
WWW: \htmladdnormallink{\protect {\tt{http://www.phys.lsu.edu/faculty/pullin}}}
{http://www.phys.lsu.edu/faculty/pullin}\\
\hfill ISSN: 1527-3431
\begin{rawhtml}
<P>
<BR><HR><P>
\end{rawhtml}
\end{flushleft}
\pagebreak
\section*{Editorial}

Just wanted to mention that the Topical Group continues to grow.
We now have 660 members, 10\% more than last year (!) and we are
now the third largest topical group. We are about half the size
of the smallest divisions of APS.

I want to encourage the readership to suggest
topics for articles in MOG. In the last few issues articles were
solicited by myself. This is not good for keeping the newsletter
balanced. Either contact the relevant correspondent or me directly.

The next newsletter is due February 1st. All issues are available in
the WWW:\\\htmladdnormallink{\protect
{\tt{http://www.phys.lsu.edu/mog}}} {http://www.phys.lsu.edu/mog}\\

The newsletter is  available for
Palm Pilots, Palm PC's and web-enabled cell phones as an
Avantgo channel. Check out 
\htmladdnormallink{\protect {\tt{http://www.avantgo.com}}}
{http://www.avantgo.com} under technology$\rightarrow$science.

A hardcopy of the newsletter is distributed free of charge to the
members of the APS Topical Group on Gravitation upon request (the
default distribution form is via the web) to the secretary of the
Topical Group.  It is considered a lack of etiquette to ask me to mail
you hard copies of the newsletter unless you have exhausted all your
resources to get your copy otherwise.  

If you have comments/questions/complaints about the newsletter email
me. Have fun.

\bigbreak
   
\hfill Jorge Pullin\vspace{-0.8cm}
\parskip=0pt
\section*{Correspondents of Matters of Gravity}
\begin{itemize}
\setlength{\itemsep}{-5pt}
\setlength{\parsep}{0pt}
\item John Friedman and Kip Thorne: Relativistic Astrophysics,
\item Raymond Laflamme: Quantum Cosmology and Related Topics
\item Gary Horowitz: Interface with Mathematical High Energy Physics and
String Theory
\item Beverly Berger: News from NSF
\item Richard Matzner: Numerical Relativity
\item Abhay Ashtekar and Ted Newman: Mathematical Relativity
\item Bernie Schutz: News From Europe
\item Lee Smolin: Quantum Gravity
\item Cliff Will: Confrontation of Theory with Experiment
\item Peter Bender: Space Experiments
\item Riley Newman: Laboratory Experiments
\item Warren Johnson: Resonant Mass Gravitational Wave Detectors
\item Stan Whitcomb: LIGO Project
\item Peter Saulson: former editor, correspondent at large.
\end{itemize}
\section*{Topical Group in Gravitation (GGR) Authorities}
Chair: John Friedman; Chair-Elect: Jim Isenberg; Vice-Chair: 
Jorge Pullin;
Secretary-Treasurer: Patrick Brady; Past Chair: Richard Price; 
Members at Large:
Bernd Bruegmann, Don Marolf, 
Gary Horowitz, Eric Adelberger, 
Ted Jacobson, Jennie Traschen. 
\parskip=10pt
\vfill
\pagebreak

\section*{\centerline {
We hear that...}}
\addtocontents{toc}{\protect\medskip}
\addtocontents{toc}{\bf GGR News:}
\addcontentsline{toc}{subsubsection}{\it  
We hear that... by Jorge Pullin}
\begin{center}
Jorge Pullin, Louisiana State University
\htmladdnormallink{pullin@phys.lsu.edu}
{mailto:pullin@phys.lsu.edu}
\end{center}

{\em Matt Choptuik, Peter Saulson and Jeff Winicour} were elected fellows of
APS.

{\em Eanna Flanagan} was elected vice-chair of the Topical Group.

{\em Sean Carroll and Bei-Lok Hu} were elected to the executive committee of the
Topical group.

{\em Warren Johnson and Bill Hamilton} were honored with the Francis Slack award
of the Southeastern Section of APS.

{\em Rodolfo Gambini} was awarded the physics prize of the Third World Academy of
Sciences (TWAS) in Trieste.

{\em Jim Hough} was honored with the Duddell Medal of the Institute of Physics (UK).

Hearty congratulations!
\vfill
\eject

\section*{\centerline {
Too many coincidences?}}
\addtocontents{toc}{\protect\medskip} \addtocontents{toc}{\bf
Research Briefs:} \addcontentsline{toc}{subsubsection}{\it 
Too many coincidences?, by Laura Mersini}
\begin{center}
Laura Mersini, Perimeter Institute
\htmladdnormallink{lmersini@perimeterinstitute.ca}
{mailto:lmersini@perimeterinstitute.ca}
\end{center}

Recent developments in precision cosmology have presented theoretical
physicists with a tantalizing picture of the universe. By a combination of
all data, there is undisputable evidence that our universe is
accelerating [1]. This means that about $70\%$ of the energy
density in the
universe is made up by a mysterious component, coined dark energy, with an
equation of state $-1.2 \leq w_X \leq -0.8$ at $68\%$ confidence
level [2].

Two fundamental questions arise in addressing the dark energy (DE)
puzzle which make this problem notoriously difficult to answer: its
magnitude $\rho_{X} \simeq 10^{-122} M_P^4$ is 122 orders less than
the expected value $M_P^4$. This is known as the fine-tuning problem;
DE domination time over matter energy density in driving the expansion
of the universe occurs around redshifts $z\simeq 0$ when the present
value of the Hubble radius is $H_0 \simeq 10^{-33} eV$. The latter is
known as the coincidence problem of DE [3]

Cosmic microwave background (CMB) measurements have proven a powerful
tool in confirming a concordance $\Lambda CDM$ picture in cosmology,
although we still lack an understanding of the origin and nature of DE
and dark matter. Together these components make for about $95\%$ of
the energy density in the universe's budget.

The $WMAP$ balloon born experiment confirmed the CMB picture of
 concordance cosmology as previously measured by $COBE$. One of the
 more surprising findings of $WMAP$ was the suppression of power at
 large angles, (low multipoles $l$), of temperature correlations
 $C_l^{TT}$ in the CMB anisotropy spectrum [4]. These findings
 can not be considered as conclusive evidence because of the
 limitations set by cosmic variance. However they are intriguing
 enough to motivate further effort in circumventing cosmic
 variance. This can be achieved by means of complimentary data like
 cosmic shear from weak lensing [5] and cross-correlations
 with the polarization spectra [6]. Analysis along these lines
 is lending support to $WMAP$ findings that indeed power is suppressed
 at low multipoles $l$. The suppressed modes correspond to
 perturbation wavelengths of the order of our present Hubble horizon
 $\lambda \simeq H_0^{-1} \simeq 10^4 Mpc, k\simeq H_0\simeq
 10^{-33}eV $.  Contrary to theoretical expectations based on the
 inflationary paradigm, not only do we have to explain the reason why
 these modes are suppressed but we also have to address why the
 suppression occurs at the DE scale, $H_0 \simeq 10^{-33} eV$. Power
 suppression at horizon sized wavelengths thus introduces a {\it second
 cosmic coincidence} to theoretical cosmology. Recall that in an
 inflationary universe perturbations produced near the end of
 inflation leave the horizon whenever their wavelength becomes larger
 than the inflationary horizon $H_i$ due to 'super-luminal'
 propagation. These modes re-enter the horizon at later times when the
 Hubble parameter once again becomes equal to their wavelength. This
 is known as the horizon crossing condition $k = a(t) H(t)$. Thus the
 largest wavelengths are the first ones to leave the horizon and the
 last ones to re-enter. Modes currently re-entering $k_0 =a_0 H_0$
 have wavelengths horizon size,which means they have been outside of
 the Hubble horizon for most of the history of the universe.  Thus
 they have not been contaminated by the internal evolution and
 nonlinearities of the cosmic fluid inside the Hubble radius. These
 modes carry the pristine information of the unknown physics which
 sets the Initial Conditions of the universe [7].

Although these cosmic coincidences associated with the two currently
observed phenomena namely, DE domination and CMB power suppression at
horizon sized wavelengths, are dominantly displayed at low energies,
for the reasons mentioned above it is reasonable to expect that they
may originate from processes occurring in the very early universe.

This is a strange world. A vacuum energy component should enhance
power of long wavelengths due to the integrated Sachs Wolf effect
(ISW). Hence we can not dismiss that the observational data seems to
point us to the existence of two cosmic coincidences at the present
Hubble radius. The bizarre picture of the universe emerging from
observational findings for these 'seemingly unrelated' cosmic
coincidences occurring at the same energy scale, may likely provide
clues of new physics.

String theory and quantum gravity are possible candidates of the
unknown physics of the early universe. There are current models in
literature that offer an explanation for the CMB power suppression, by
having the Initial Conditions set within the framework of string
theory [7,8] loop quantum gravity [9]  or an
unknown hard cutoff [10]. There is also an ongoing search for a
possible $UV/IR$ mixing of gravitational scales [11]. However a
theoretical model that would successfully accommodate all observed
cosmic coincidences around the scale $H_0 \simeq 10^{-33}eV$ is yet to
be found.

Perhaps, as the data is suggesting, there is something special about
our present Hubble scale. It might be a fundamental scale of very low
energy physics. Or perhaps a new scale of low energies derived from a
fundamental scale of high energy physics through a possible
$UV/IR$. This radical possibility is not yet realized in a concrete
model.

At the moment, our theoretical knowledge of the relation between
beauty and a strange world still lies in the realm of speculations
while pushing forward the discovery of new physics.

{\bf References:}

\parskip=0pt [1] Knop, R.A.et al.(2003),
\htmladdnormallink{astro-ph/0309368}
{http://arXiv.org/abs/astro-ph/0309368}; Spergel, D.N. et al.(2003),
ApJS148, 175; Verde, L. et al., MNRAS335, 432.

[2] Melchiorri, A., Mersini, L., Odman, C., Trodden, M. (2003),
Phys.Rev.D68,43509, \htmladdnormallink{astro-ph/0211522}
{http://arXiv.org/abs/astro-ph/0211522}.

[3] Carroll, M.S., \htmladdnormallink{astro-ph/0310342}
{http://arXiv.org/abs/astro-ph/0310342}, (2003), and references
therein.

[4] Bennett,et al., \htmladdnormallink{astro-ph/0302207}
{http://arXiv.org/abs/astro-ph/0302207} Astrophys.J.Suppl. 148 (2003);
Hinshaw, et al., \htmladdnormallink{astro-ph/0302217}
{http://arXiv.org/abs/astro-ph/0302217}, Astrophys.J.Suppl. 148
(2003); Tegmark et al, \htmladdnormallink{astro-ph/0302496}
{http://arXiv.org/abs/astro-ph/0302496}; Melchiorri, A.,
\htmladdnormallink{hep-ph/0311319}{http://arXiv.org/abs/hep-ph/0311319}, (2003).

[5] Kesden, M., Kamionkowski, M., Cooray, A.,
\htmladdnormallink{astro-ph/0306597}
{http://arXiv.org/abs/astro-ph/0306597}, (2003).

[6] Dore, O., Holder, G.P., Loeb, A.,
\htmladdnormallink{astro-ph/0309281}
{http://arXiv.org/abs/astro-ph/0309281}, (2003).

[7] Bastero-Gil,M., Freese, K., Mersini-Houghton,L.,  
\htmladdnormallink{hep-ph/0306289}
{http://arXiv.org/abs/hep-ph/0306289}, (2003), and references therein.

[8]  Dvali, G., Kachru, S., 
\htmladdnormallink{hep-th/0309095}
{http://arXiv.org/abs/hep-th/0309095}, (2003).

[9] Tsujikawa,S., Singh, P., Maartens, R.,
\htmladdnormallink{astro-ph/0311015}
{http://arXiv.org/abs/astro-ph/0311015}, (2003).

[10] Contaldi, C.R., Peloso, M., Kofman, L., Linde, A., JCAP
0307:002,(2003), \htmladdnormallink{astro-ph/0303636}
{http://arXiv.org/abs/astro-ph/0303636}.

[11] Banks, T., Fischler, W., \htmladdnormallink{astro-ph/0307459}
{http://arXiv.org/abs/astro-ph/0307459} and references therein; also
see Banks, T., 
\htmladdnormallink{hep-th/0310288}
{http://arXiv.org/abs/hep-th/0310288}, (2003).
\parskip=10pt
\vfill
\eject
\section*{\centerline {
The Quest for a Realistic Cosmology}\\
\centerline{in the Landscape of String Theory}}
\addcontentsline{toc}{subsubsection}{\it  
The Quest for a Realistic Cosmology in String Theory, by Andrew Chamblin}
\begin{center}
Andrew Chamblin, University of Louisville
\htmladdnormallink{chamblin@prancer.physics.louisville.edu}
{mailto:chamblin@prancer.physics.louisville.edu}
\end{center}

Recent astronomical observations [1,2,3,4] would appear to indicate
that the universe is accelerating.  Assuming that these observations
have been correctly interpreted, then it is clear that physicists
today are faced with a number of mysteries which to date have defied
any elegant and straightforward explanation.  First of all, there is
the obvious question: What is the nature of this mysterious `dark
energy' which is driving the expansion?  Evidently this vacuum energy
is exactly isotropic and homogeneous at the present time - but what is
it?  In addition to this question, there is the legendary
`cosmological constant problem': Whatever this dark energy is, why is
it so incredibly small?  Observationally, the dark energy density is
120 orders of magnitude smaller than the energy density associated
with the Planck scale - the obvious cut off.  Furthermore, the
standard model of cosmology posits that very early on the universe
experienced a period of inflation: A brief period of very rapid
acceleration, during which the Hubble constant was about 52 orders of
magnitude larger than the value observed today.  How could the
cosmological constant have been so large then, and so small now?
Finally, there is the `coincidence problem': Why is the energy density
of matter nearly equal to the dark energy density today?  Considering
all of these problems at once can be a humbling experience: It is
clear that we are presently unable to explain several of the most
basic experimental facts about this universe.

String theory is much vaunted as a fully consistent quantum theory of gravity.
If this is the case, then we would expect string theory to tell us
{\it something} about the acceleration of the universe.  Ideally, string
theory would provide clear mechanisms which resolve all of the above
mentioned problems.

Remarkably, it is very difficult to get accelerating
solutions directly through standard compactification techniques
of the low-energy limit of string theory.  More precisely, the
different string theories are related to each other through dualities,
special symmetries which ultimately involve the mysterious quantum
M-theory in eleven dimensions.  All of these theories have as a low-
energy limit some supergravity theory: A {\it classical} theory 
consisting of gravity coupled to other fields.  These theories should
be thought of as special limits of some underlying, quantum M-theory.
We do not know what the entire moduli space of this theory looks like, but
we know what it looks like at these special limit points.  A major triumph
for string theory would be to show precisely `where' in the M-theory
moduli space there exist solutions which actually look like our universe.
For example, it would be nice if we could recover a realistic M-theory
cosmology beginning with one of the supergravity theories.  However, there is 
a `no-go theorem', due to Gibbons [5], Maldacena and Nu\~nez [6],
and also de Wit, Dass and Smit [7], 
which basically asserts that if you compactify any
string-derived supergravity on a smooth compact internal space, then
you will never get de Sitter space.  Since the universe is evidently
both past and future de Sitter (albeit with vastly differing vacuum 
energies), this would seem to be a problem.

However, there are various ways around this particular no-go result.
The theorem assumes time independence of the internal space, and so
one may search for time-dependent solutions.  Following this
intuition, Townsend, Wohlfarth and others [8,9,10] 
have constructed a variety of time-dependent
compactifications which describe a period of acceleration.  The basic
idea is simple: The internal space is described by certain scalar
fields known as `moduli'.  These moduli describe the size, shape and
other basic properties of the internal space.  These moduli typically
have exponential potentials, with the property that as you flow to the
minimum of the potential the universe decompactifies (i.e., for a
given scalar field $\phi$, $V(\phi) ~{\sim}~ e^{-\phi}$).  One can now
imagine `bouncing' the universe off of this potential: The universe
comes in from a period of being decompactified, and rolls up the
exponential potential during the process of compactification.  At some
point there will have to be a `turnaround' point, where the universe
stops compactifying and reenters a decompactification stage.  At the
turnaround, there is little kinetic energy for the moduli, and so all
of the energy is dominated by the potential term, which can then act
as a cosmological constant.  Problems with this approach include the
fact that it is difficult to get a very long period of inflation,
unless one uses many moduli [11].  Furthermore, if the size
of the extra dimensions vary, then there will be variations in
Newton's constant and the fine structure constant.  Strong
experimental bounds on such variation place tight constraints on these
models.

Another way to get around the no-go result involves beginning with
rather exotic supergravities which may not necessarily have anything
to do with string or M-theory.  For example, Hull has championed the
viewpoint that we may wish to consider supergravity theories with
extra dimensions of time [12].  These supergravity actions
come from perfectly well-defined superalgebras, and compactification
of these theories can give de Sitter spacetime in any dimension.  One
drawback is that these theories typically contain {\it ghosts}: Gauge
fields which have the wrong sign for the kinetic term.  Furthermore
there are the usual problems with causality: If you have two or more
dimensions of time, then there exist closed timelike curves through
every point.  While the extra dimensions of time can be eliminated by
applying certain duality transformations which yield another theory,
one is often still left with ghosts.

One may also choose to `compactify' a string-derived supergravity
on a {\it non-compact} space [13].  This gets around
the no-go result because the internal space is non-compact.  This may sound
counterintuitive, but actually it is a well-defined procedure known
as 'consistent truncation'.  To perform a consistent truncation,
one writes the full higher-dimensional space as a product (or in general
warped product) of the non-compact directions and some space X.  One
then constructs a theory on X, with the property that any solution of
that theory corresponds to a solution of the theory in the higher dimensions,
and vice-versa.  In this way one can obtain solutions where X is
isometric to de Sitter.  One obvious problem with this approach is that it is not
clear how one should interpret the large extra dimension.

A related but differing approach involves using the scale invariance
of eleven-dimensional supergravity, which is the low-energy limit of
M-theory.  The equations of motion of eleven-dimensional supergravity
admit a scale invariance, whereby rescaling the field content in a
certain way simply rescales the action by an overall power of the
scale parameter.  Instead of compactifying the theory on a circle
using `conventional' Kaluza-Klein boundary conditions (where the
fields are periodic), one can use the scale symmetry to allow fields
to be rescaled around the circle.  Upon reduction to ten dimensions
one obtains a new massive supergravity theory, which has the property
that de Sitter space is the ground state.  Intuitively, the apparent
expansion of the universe is really an effect generated by the
rescaling of the metric.  This theory was first introduced by Howe,
Lambert and West [14] and was obtained through consistent
truncation by others [15].  It was further studied by this
author and Lambert [16,17] where we dubbed the
theory `MM-theory', for modified or massive M-theory.  The main
problem with MM-theory is that the scale invariance is an anomalous
symmetry: Higher derivative corrections to the supergravity Lagrangian
manifestly break the symmetry.  Since the theory is anomalous in the
ultraviolet, the only way it can make sense is if scale invariance is
realized deep in the infrared.  In this picture were correct, then the
cosmological constant itself would be an infrared effect.

Ultimately, all of these classical approaches to cosmology seem a bit
contrived: In order to get around the no-go theorem, one is forced to
make rather unusual or unnatural assumptions.  But of course, all of
these expeditions are only probing the {\it classical} borders of the
full landscape of string theory.  The world is not classical: There is
an underlying quantum reality, and we need to better understand the
classical to quantum phase transition within the context of cosmology.
Could it be that if we simply venture into the quantum wilderness of
the string theory landscape, we will find a realistic cosmology?

In fact, it is the case that quantum effects seem to lead in the right direction.
In a recent paper, by Kachru, Kallosh, Linde and Trivedi (KKLT) [18]
it was shown that if you carefully consider certain instanton corrections, you can
construct solutions of string theory which exhibit a small, positive cosmological
constant.  Their example is an example of a `flux compactification' - crudely,
a compactification in which certain fluxes are turned on.  Typically, certain branes
are the `sources' for a given flux.  For example, just as the electron is the source
for $F_{ab}$ (a two-form), so a membrane can act as the `electric' source for a 
four-form flux.  In four dimensions, the equations of motion for a four-form will
tell you that the form is locally just covariantly constant:  The term 
$(F_{4})^2$ in the action will thus `look' like a cosmological constant.  Membranes
in such scenarios are thus surfaces across which the effective cosmological constant
can jump.  Neutralization of the cosmological constant through membrane nucleation was
first studied by Brown and Teitelboim [19], and has been further explored in the
context of string theory by others [20,21].

In the KKLT construction, the authors begin by compactifying six
dimensions of space on a Calabi-Yau manifold - a complex manifold
which has a special holonomy that leaves minimal supersymmetry ($N =
1$) in the effective four-dimensional theory obtained through the
compactification.  Certain background fluxes are turned on throughout
the construction, and in the effective four-dimensional theory the
spacetime is initially anti-de Sitter (adS).  A Calabi-Yau manifold
has certain moduli associated with the fact that it admits a complex
structure, and these moduli need to be fixed.  KKLT show that this is
possible by arguing that quantum effects modify the superpotential
[22] in such a way that they are able to explicitly demonstrate the
existence of supersymmetric adS vacua with fixed complex and Kahler
moduli.

Finally, and crucially, KKLT add in certain branes, known as (`anti')
D3-branes.  These branes have the effect of `lifting' the stable (and
supersymmetric) adS vacuum to a {\it de Sitter} (dS) vacuum.  By fine
tuning various things, the authors are able to argue that the
resulting dS vacuum can even have a very small cosmological constant
\footnote{It is worth pointing out that there has been some
non-trivial criticism of this construction.  Put crudely, the
construction is entirely at the level of effective field theory, and
it is unclear that it really can be embedded into a full `stringy'
setup.  Since this is a technical point beyond the scope of this
review, I will have nothing more to say about it here.}.  Furthermore,
the inclusion of the three-branes breaks supersymmetry, and so it
would seem that supersymmetry breaking and a positive cosmological
constant always go `hand-in-hand' in these constructions.  Finally,
the de Sitter vacua are always metastable in these models, i.e., they
are false vacua and therefore have some lifetime.  In particular, KKLT
argue that these vacua are resonances which can decay faster than the
timescale for the Poincare recurrences which have bothered some people
[23].

Now, the KKLT model is but a very special case of a huge class of more
general compactifications.  One can imagine solutions where only four
dimensions are compactified, or indeed where none of the dimensions
are compactified and the universe exhibits the full eleven dimensions
of M-theory.  One could imagine that other fluxes are turned on, or
that no fluxes are turned on.  If we think of the string theory
landscape as a huge potential or functional which varies depending on
all of the different possible moduli, then it is clear that the
quantum wilderness of string theory is a vast, higher dimensional
cornucopia of moutaintops, valleys and precipices.  The moduli space
of supersymmetric vacua is rather like a vast plain extending up to
the mountains - one may move continuously between different vacua by
varying certain moduli.  Accelerating cosmologies correspond to
isolated valleys which sit up between the mountain peaks and passes
(i.e., one might imagine equating the magnitude of the dark energy
with the altitude of the valley).  For whatever reason, we live in a
universe where four spacetime dimensions are compactified, and our
`altitude' is just {\it barely} above sea level.

Of course, when one starts to think of the universe in these terms, it
can have a profound impact on one's expectations and outlook.  It
starts to look like many things - the masses of the elementary
particles, the values of the couplings, the value of the cosmological
constant - are probably just accidents, random numbers that will never
be calculated from first principles using string theory.  But it is a
short journey from this philosophy to that house of ill-repute known
as the Anthropic Principle \footnote{To quote Lenny Susskind: `We live
where we can live.', (New York Times, Sept. 2, 2003)}.  For this
reason, various people have begun to `count' [24] all of the different
discrete valleys in the string theory landscape.  After all, it may be
that many of the vacua look somewhat like our universe - and even if
only about 1\% look like home, we will have still learned something
about our world.

In summary, cosmology is now a science based on high precision
measurements which are yielding very detailed information about the
large-scale structure of the universe.  For some time it was unclear
that string theory could consistently explain the observed
acceleration of the universe.  This situation has now been rectified,
and there is now a realization that there are likely many metastable,
de Sitter like vacua in string theory.  These are just the first
tentative steps towards a fully realistic string cosmology, and the
years ahead will no doubt bring even more twists, turns and surprises.

I thank N. Lambert and H. Reall for discussions, and the Kavli
Institute for Theoretical Physics for hospitality while this work was
completed.

[1] A.G. Riess {\it et al}, Astron. J. 116 (1998) 1009-1038;
\htmladdnormallink{astro-ph/9805201}
{http://arXiv.org/abs/astro-ph/9805201}.

[2] S. Perlmutter {\it et al}, Astrophys. J. 517 (1999) 565-586;
\htmladdnormallink{astro-ph/9812133}
{http://arXiv.org/abs/astro-ph/9812133}.

[3] D. Spergel {\it et al}, Astrophys. J.Suppl. 148 (2003) 175;
\htmladdnormallink{astro-ph/0302209}
{http://arXiv.org/abs/astro-ph/0302209}.

[4] L. Verde {\it et al}, Mon. Not. Roy. Astron. Soc. 335 (2002) 432;
\htmladdnormallink{astro-ph/0112161}
{http://arXiv.org/abs/astro-ph/0112161}.

[5] G. Gibbons, in GIFT Seminar on supersymmetry, supergravity and related topics,
edited by F. del Aguila, J. de Ascarraga and L. Ibanez, World Scientific (1984).

[6] Juan Maldacena and Carlos Nu\~nez, Int. J. Mod. Phys. A16 (2001)
822-855; 
\htmladdnormallink{hep-th/0007018}
{http://arXiv.org/abs/hep-th/0007018}.

[7] B.~de Wit, D.~J.~Smit and N.~D.~Hari Dass,
Nucl.\ Phys.\ B {\bf 283}, 165 (1987).

[8] P. Townsend and M. Wohlfarth, Phys. Rev. Lett. 91 (2003) 061302;

\htmladdnormallink{hep-th/0303097}
{http://arXiv.org/abs/hep-th/0303097}.  M. Wohlfarth, 
\htmladdnormallink{hep-th/0307179}
{http://arXiv.org/abs/hep-th/0307179}.

[9] F. Darabi, Class. Quant. Grav. 20 (2003) 3385-3402; 
\htmladdnormallink{gr-qc/0301075}
{http://arXiv.org/abs/gr-qc/0301075}.

[10] R. Emparan and J. Garriga, JHEP 0305 (2003) 028; 
\htmladdnormallink{hep-th/0304124}
{http://arXiv.org/abs/hep-th/0304124}.

[11] Chiang-Mei Chen, Pei-Ming Ho, Ishwaree P. Neupane, John E. Wang, JHEP 0307 (2003) 017; 
\htmladdnormallink{hep-th/0304177}
{http://arXiv.org/abs/hep-th/0304177}.

[12] C.M. Hull, 
JHEP 9807 (1998) 021; 
\htmladdnormallink{hep-th/9806146}
{http://arXiv.org/abs/hep-th/9806146}.

[13] G.W. Gibbons and C.M. Hull, 
\htmladdnormallink{hep-th/0111072}
{http://arXiv.org/abs/hep-th/0111072}.

[14] P.S. Howe, N.D. Lambert and P.C. West, 
Phys. Lett. B416 (1998) 303-308; 
\htmladdnormallink{hep-th/9707139}
{http://arXiv.org/abs/hep-th/9707139}.

[15] I.V. Lavrinenko, H. Lu and C.N. Pope, 
Class. Quant. Grav. 15 (1998) 2239-2256; 
\htmladdnormallink{hep-th/9710243}
{http://arXiv.org/abs/hep-th/9710243}.

[16] A. Chamblin and N. Lambert,  Phys. Lett. B508 (2001) 369-374; 
\htmladdnormallink{hep-th/0102159}
{http://arXiv.org/abs/hep-th/0102159}.

[17] A. Chamblin and N. Lambert, 
Phys. Rev. D65 (2002) 066002; 
\htmladdnormallink{hep-th/0107031}
{http://arXiv.org/abs/hep-th/0107031}.

[18] Shamit Kachru, Renata Kallosh, Andrei Linde and Sandip
P. Trivedi, Phys. Rev. D68
(2003) 046005; 
\htmladdnormallink{hep-th/0301240}
{http://arXiv.org/abs/hep-th/0301240}.

[19] J.C. Brown and C. Teitelboim, Nucl. Phys. B297: 787-836, 1988.

[20] Raphael Bousso and Joseph Polchinski, JHEP 0006 (2000) 006; 
\htmladdnormallink{hep-th/0004134}
{http://arXiv.org/abs/hep-th/0004134}.

[21] Jonathan L. Feng, John March-Russell, Savdeep Sethi and Frank
Wilczek, Nucl. Phys. B602 (2001) 307-328; 
\htmladdnormallink{hep-th/0005276}
{http://arXiv.org/abs/hep-th/0005276}.

[22]  S. Gukov, C. Vafa and E. Witten, 
Nucl. Phys. B584 (2000) 69-108; Erratum-ibid. B608 (2001) 477-478; 
\htmladdnormallink{hep-th/9906070}
{http://arXiv.org/abs/hep-th/9906070}.

[23]  L. Susskind,  
\htmladdnormallink{hep-th/0302219}
{http://arXiv.org/abs/hep-th/0302219}.

[24]  M.R. Douglas,  JHEP 0305 (2003) 046; 
\htmladdnormallink{hep-th/0303194}
{http://arXiv.org/abs/hep-th/0303194}.

\vfill
\eject

\section*{\centerline {{\bf SFB{$/$}TR 7:  Gravitational
 Wave Astronomy}}}
\addcontentsline{toc}{subsubsection}{\it
SFB/TR 7, by A. Gopakumar \& D. Petroff }
\begin{center}
A. Gopakumar \& D. Petroff, SFB/TR 7,
\htmladdnormallink{gopu,petroff@tpi.uni-jena.de}
{mailto:petroff@tpi.uni-jena.de}
\end{center}

   The first meeting of researchers participating in a new German
initiative on gravitational wave astronomy took place in T\"ubingen
from October 9--10, 2003.  The ambitious project entitled
`Gravitational Wave Astronomy' and referred to as SFB/TR~7
(Sonderforschungsbereich/Transregio 7) is being funded by the DFG
(Deutsche Forschungsgemeinschaft), for an initial period of four years
(2003--2006) and with the possibility of extensions for up to twelve
years.  SFB/TR~7 brings together more than 50 scientists as well as
numerous Diploma and Ph.D. students from five German academic
institutions, to tackle various issues related to the realization of
gravitational wave astronomy in the near future. Opening a new window
to the universe by observing gravitational waves requires close
interaction between experimentalists and theoreticians working on
various aspects of gravitational radiation. It is highly desirable
that experimentalists involved in the design and construction of
gravitational wave detectors associate closely with theoreticians
providing detailed information about the nature and abundancy of
expected sources and their signals, as well as with those who analyse
the raw detector data and develop new analysis strategies. It is to
achieve the above objectives that the DFG is supporting experimental
and theoretical physicists, astrophysicists and mathematicians from
Universities in Hanover, Jena and T\"ubingen as well as Max-Planck
Institutes for Astrophysics in Garching and Gravitational Physics in
Golm.

   Prof.\ G.\ Neugebauer (Jena) is the speaker of SFB/TR 7 and the
executive board is also comprised of Profs.\ K.\ Danzmann (Hanover),
W.\ Kley (T\"ubingen), E.\ M\"uller (Garching), G.\ Sch\"afer (Jena)
and B.\ Schutz (Golm).  The project is subdivided into three sections
dealing with the analysis of the gravitational field equations, the
structure and dynamics of compact objects and the detection of
gravitational waves.  Each of these sections consists of several
working groups, which tackle specific issues relevant to the parent
section as well as to gravitational wave astronomy in a broader sense.

During the meeting, each working group presented a brief status report
of its work. Junior scientists were encouraged to assume the
responsibility of preparing and delivering the talks, which helped to
lend the conference a relaxed and open atmosphere. The three working
groups analysing the structure of the field equations relevant to
numerical simulations were the first to present their projects. This
portion of the meeting was made up of the six talks listed below:
$\bullet$ Vacuum Initial Data with Trapped Surfaces $\bullet$ A
Program for the Numerical Treatment of Radiating Systems $\bullet$
Gravitational Radiation from Distorted Black Holes $\bullet$ Initial
Data for the Conformal Einstein Equations $\bullet$ A Skeleton
Solution of the Einstein Field Equations and $\bullet$ A Minimal
No-Radiation Approximation to the Einstein Field Equations.

The second, and largest section in SFB/TR 7, is made up of six working
groups, which presented ten talks related to their respective research
interests. These can be broadly classified, according to the
underlying astrophysical scenarios, into four categories. The first
deals with the structure of solitary compact objects, the second with
their dynamics and the third with the collapse of relativistic
objects. The fourth category focuses on binary dynamics within the
post-Newtonian and numerical relativity frameworks. The titles of
these talks (in the order presented at the workshop) are $\bullet$ An
Updated Version of a Computer Program for the Calculation of Rotating
Neutron Stars and Specific Applications $\bullet$ Oscillation Modes of
Rotating Neutron Stars $\bullet$ New Methods for Gravitational
Collapse to Neutron Stars and Black Holes $\bullet$ Gravitational
Collapse of Rotating Neutron Stars $\bullet$ Cylindrical Collapse
$\bullet$ Binary Dynamics of Spinning Compact Objects $\bullet$
Gravitational Waves from Binary Systems with Oscillating Dust Discs as
Components $\bullet$ Evolutions in 3D Numerical Relativity using Fixed
Mesh Refinement $\bullet$ Binary Black Hole Evolutions from Innermost
Quasi-Circular Orbits and $\bullet$ Merging Neutron Star Binaries --
Results and Future Plans.

On the second day, the experimentalists and data analysts making up
the third section of our SFB, briefed the participants on their
research progress. The experimentalists gave talks on both currently
implemented technologies, e.g.\ in GEO600, and on possible future
methodologies with the titles $\bullet$ High Resolution Interferometer
Concepts Based on Reflective Optical Components, $\bullet$ Low Loss
Gratings for Gravitational Interferometry, Design Considerations and
Fabrication and $\bullet$ Cryogenic Q-factor Measurement of Optical
Substrates. The session on data analysis consisted of a detailed
review and a discussion session on the iterative design of the
sensitivity curve for future gravitational wave detectors.

Those interested in learning more about SFB/TR~7 are referred to the
website 

\htmladdnormallink{
http://www.tpi.uni-jena.de/\~{}sfb/index.html}
{http://www.tpi.uni-jena.de/\~sfb/index.html}.

\vfill
\eject

\section*{\centerline{The mock LISA data archive}}
\addcontentsline{toc}{subsubsection}{\it
The mock LISA data archive, by John Baker}
\begin{center}
John Baker, Goddard
\htmladdnormallink{jbaker@milkyway.gsfc.nasa.gov}
{mailto:jbaker@milkyway.gsfc.nasa.gov}
\end{center}

The LISA gravitational wave observatory is expected to observe a broad
variety of astrophysical phenomena, and may discern detailed
characteristics of thousands of individual astronomical systems.  But
realizing these goals will require careful resolution of a variety of
data analysis challenges.  The Mock LISA Data Archive (MLDA), hosted
at the AstroGravS website 
(\htmladdnormallink{http://astrogravs.nasa.gov}{http://astrogravs.nasa.gov}) 
is a resource
for researchers interested in developing algorithms for analyzing LISA
data.  Our goal is to facilitate collaboration and to encourage
additional researchers to become involved in the LISA data analysis
community.
 
The archive contains simulated LISA output data representing the
instrument's response to several classes of gravitational wave
sources. State-of-the-art model waveforms from these sources have been
run through a simulation of LISA's response.  The MLDA currently
contains four source classes: Supermassive Black Hole Binaries;
Extreme Mass Ratio Captures; Galactic Binaries; and realizations of
the Galactic Background. The present simulations of LISA's response
have been produced with the publicly available LISA Simulator
(\htmladdnormallink{http://www.physics.montana.edu/lisa/}
{http://www.physics.montana.edu/lisa/}). New source data and more
advanced response simulations will be added as they become
available. Contributions to the MLDA are most welcome.

The MLDA provides for the development of techniques to tackle each
source class individually, or the signals can be combined to produce a
more realistic test of a data analysis procedure.  A fixed signal
stream can be combined with multiple noise realizations, which is
useful for Monte-Carlo studies of algorithm performance.

One of the goals of the MLDA is to provide a common ground for
comparing different approaches to LISA data analysis. Each simulated
data stream comes with reference files containing a description of the
sources that have been modeled. Thus, the MLDA can serve as a training
ground for the LISA data analysis community.

At some time in the future, prior to the launch of the LISA observatory,
the MLDA hopes to host a Mock LISA Data Challenge to see which algorithms
perform best in a blind test using realistic multi-source Mock Data sets.

The MLDA is designed to be a community resource, with input from the
entire LISA community. We invite interested researchers to become involved
in LISA data analysis, to utilize the archive, and to participate in its
further development. Please let us know what you think, and how you would
like to be involved.

MLDA Steering Committee:

Neil Cornish, Convener, Montana State University,
John Baker, GSFC,
Matt Benacquista, Montana State University--Billings,
Joan Centrella, GSFC,
Scott Hughes, MIT,
Shane Larson, Caltech.

\vfill
\eject

\section*{\centerline {
Second Gravitational Wave Phenomenology Workshop}}
\addtocontents{toc}{\protect\medskip}
\addtocontents{toc}{\bf Conference reports:}
\addcontentsline{toc}{subsubsection}{\it  
Second Gravitational Wave Phenomenology Workshop, by Neil Cornish}
\begin{center}
Neil Cornish, Montana State University
\htmladdnormallink{cornish@physics.montana.edu}
{mailto:cornish@physics.montana.edu}
\end{center}

It is likely that Sam Finn had to twist some arms to get a good
turnout of astrophysicists at the First Gravitational Wave
Phenomenology Workshop in November 2001, but two years later, with the
scent of data in the air, he would have had a hard time keeping them
out of the second workshop, which took place at Penn State on November
6-8, 2003.

The goal of the workshop was to foster dialog between observers,
astrophysical source modelers, and data analysts, on the question
of how gravitational
wave observations could be used to ``inform our understanding of the
cosmos'' (to quote the workshop website,
\htmladdnormallink{http://cgwp.gravity.psu.edu/events/GWPW03/}
{http://cgwp.gravity.psu.edu/events/GWPW03/}). The workshop was designed
to be discussion heavy, with short talks providing the starting point
for longer discussions.

The meeting began with an update on the three types of gravitational wave
detectors currently in operation, and the implications that these
observations have on astrophysical models. Stan Whitcomb described the
various ground based Laser Interferometers; Giovanni Prodi talked about
Acoustic (Bar) Detectors; and Andrea Lommen discussed Pulsar Timing Arrays.
A common theme was that even null detections can be used to place
bounds on astrophysical processes, and while the current upper limits
are not that strong in most cases, we are already doing gravitational
wave astronomy. Andrea gave one example where Pulsar Timing had been able to
rule out the suggestion that the Radio Galaxy 3C66B harbors a supermassive
black hole binary of $5.4 \times 10^{10} \, M_{\odot}$. Another common
theme in the detector session was optimism for the future. Stan boldly
predicted that LIGO and VIRGO would reach their design sensitivities
in 2004, Giovanni saw a bright future for dual resonant bars, and
Andrea foresaw Pulsar Timing Arrays being able to detect the gravitational
wave background produced by supermassive black hole binaries.
The impact of the observations on our understanding of Neutron Star Physics,
Stellar Populations and Burst Sources was then addressed by Ben Owen,
Vicky Kalogera and Andrew MacFadyen. Ben outlined how targeted searches
for periodic signals from know Pulsars could be used to constrain the
material properties of a Neutron Star's crust. Vicky emphasized that
large uncertainties in the theoretical models mean that any observational
input - upper bounds or a few direct detections - would strongly
constrain certain stellar population models. As an illustration of this
point, Vicky showed how the recent discovery of a third binary pulsar
in our galaxy has lead to a six-fold increase in the predicted Neutron Star
Inspiral event rate for LIGO. In the same vein, Andrew explained how
uncertainties in the modeling of Gamma Ray bursts would be reduced by
co-ordinated gravitational and electromagnetic observations, even if
no gravitational wave counterparts were found.

The second day of the workshop was devoted to gravitational wave source
astrophysics and source modeling. Brad Hansen reviewed the evidence for
Intermediate Mass Black Holes (IMBH) and brought up the interesting possibility
that an IMBH might be responsible for dragging the observed population of
young, massive stars into the lair of the supermassive black hole at the
galactic center. Christian Cardall and Adam Burrows described
supernova modeling and what the gravitational wave signatures might
look like. They had good news for gravitational wave astronomers - supernova
simulations fizzle unless there is asymmetry. Adam showed some possible
waveforms with very interesting ringing structure. Milos Milosavljevic
took on the ``last parsec problem'' and concluded that it probably
wasn't much of a problem - a range of physical
mechanisms could succeed in getting black hole binaries close enough to
evolve under radiation reaction. Leor Barak brought us up to date on
the impressive progress that has been made in solving the self-force
problem, and how the new calculation schemes were being used to model
extreme mass ratio inspiral. Deirdre Shoemaker and Masaru Shibata
led the discussion on numerical relativity, and its application in
the areas of black hole and neutron star binary evolution. Both
conceded that significant problems remained to be solved, especially
when black holes enter the picture, but they also showed how partial
results were better than no results. Shibata drew an analogy between
numerical relativity and the operation of the laser interferometers.
The system is up and running, and while the performance is below
the design goals, they are collecting data and learning as they go.

The final day had us looking to the future, and the promise of advanced
detectors. Sam Finn discussed second generation ground based laser
interferometers, and emphasized how astrophysical considerations will
play a role in making design choices. Eugenio Coccia described how future
acoustic detectors, with their improved sensitivities and wider bandpasses,
can provide coverage of the very high frequency portion of the spectrum.
The final speaker, Neil Cornish, got a little carried away with the
``advanced detector'' theme, and after a brief description of the
Laser Interferometer Space Antenna (LISA) and its potential impact, he went on
to describe two post-LISA missions. These follow-on missions use multiple
detectors for improved angular resolution, and in the case of the
Big Bang Observatory, co-aligned interferometers for detecting the
Cosmic Gravitational Wave Background.

A proper account of the workshop would report on the discussions that
followed the talks, as these discussions involved all the workshop
participants, and took us well beyond the formal presentations. But
lacking extensive notes, I'll have to skip the best part and simply
encourage people to attend the third meeting in the series.

\parskip=5pt
\vfill\eject
\section*{\centerline{ 
``Apples with Apples II'':
Second Workshop on Formulations}\\ 
\centerline{of Einstein's Equations
for Numerical
Relativity}}
\addcontentsline{toc}{subsubsection}{\it  
Apples with apples II, by Miguel Alcubierre}
\begin{center}
Miguel Alcubierre, UNAM, Mexico
\htmladdnormallink{malcubi@nuclecu.unam.mx}
{mailto:malcubi@nuclecu.unam.mx}
\end{center}

Last December 1-11 saw the second ``Apples with Apples'' workshop on
formulations of the Einstein equations for numerical relativity.  As
in the case of the first workshop, the venue was the Institute of
Nuclear Sciences (ICN) of the National University of Mexico (UNAM), in
Mexico city.  This second workshop was attended by some 40 people from
Mexico, the U.S., Europe and Japan, that is, considerably more than
the 25 who attended the first workshop.  The workshops are getting
more popular!

As before, the purpose of the workshop was to gather a group of
experts in the recent developments of the different formulations of
the Einstein equations and their applications to numerical relativity.
In particular, these workshops have as their main focus both the
sharing of ideas as well as the direct comparison of the results of
actual numerical simulations, in the hope of learning what makes one
formulation better suited for numerical work than others.  The
workshops have a very open and relaxed format, extended over a two
week period, to allow time for informal discussions and work sessions
(and I mean really work sitting in front of the laptop).  Of course,
eating mexican food and drinking tequila are also a part of it, as our
boat trip in the canals of Xochimilco can show.

There where over 25 talks this time, and it would take too much space
to mention them all here (but have a look at the web page
\htmladdnormallink{http://www.appleswithapples.org}
{http://www.appleswithapples.org}, where most of the talks can be found
as PS or PDF files).  In my own personal opinion, the highlights
included Jeff Winicour's introductory talk describing the aims of the
Apples with Apples meetings, Denis Pollney's talk on test-suites for
numerical relativity, Carsten Gundlach's talk on well posed BSSN,
Scott Hawleys' talk on constrained evolution, Hisaaki Shinkai and Gen
Yoneda's talk on constraint propagation analysis, and Manuel Tiglio's
talk on dynamical control of constraint violation.  Of special
interest where a series of talks on well posed boundary condition,
from Jeff Winicour and Bela Szilagyi's work with the Abigel code, to
Mihaela Chirvasa's talk on absorbing boundary conditions, and of
course Olivier Sarbach's talk on constraint preserving boundary
conditions.  Results from the conformal field equation approach were
also presented by Sascha Husa and Christiane Lechner.  A particular
highlight was the participation of Nina Jansen and her passionate
discussion of tests performed on different systems following the paper
coming out of the first Apples with Apples meeting (
\htmladdnormallink{gr-qc/0305023}{http://arXiv.org/abs/gr-qc/0305023}).
Nina's participation helped us focus on which of those tests are
really useful, and what direction should be taken to suggest new
tests.  At the end of the meeting, an agreement was reached for the
several groups represented to test their codes following the test-suite
suggested in \htmladdnormallink{gr-qc/0305023}{http://arXiv.org/abs/gr-qc/0305023}.

If anything, the meeting helped us realize that the numerical
community is finally addressing the issues of well posedness of
evolution systems, gauge conditions, and boundary conditions head on.
Mysteries remain as to why some well posed formulations perform well
numerically and other do not, but significant progress has certainly
been made.

And now, to the future.  Oscar Reula has graciously offered to host
``Apples with Apples III'' in Cordoba, Argentina.  Final dates are
still not clear, but the meeting will probably take place in late 2004
or early 2005.  Winter in the U.S. and Europe, but summer down below
the equator.  See you in Argentina!

\vfill
\eject

\parskip=10pt

\section*{\centerline {
3 Conferences for 30 years of Gravity at UNAM}}
\addcontentsline{toc}{subsubsection}{\it 3 Conferences for 30
years of Gravity at UNAM, by Alejandro Corichi}
\begin{center}
Alejandro Corichi, National University of Mexico (UNAM)
\htmladdnormallink{corichi@nuclecu.unam.mx}
{mailto:corichi@nuclecu.unam.mx}
\end{center}

The Department of Gravitation and Field Theory of the Institute
for Nuclear Sciences at the National University in M\'exico
(ICN-UNAM), as part of the celebrations commemorating its 30 years
of existence, was  host to three conferences the first week of
February. The first one was a 3-day discussion workshop on Loop
Quantum Gravity (LQG) with participants coming from Canada, the
USA and M\'exico. The second event was a one-day celebration for
Mike Ryan's 60th birthday, and finally a two-day celebration for
Professor Marcos Rosenbaum, for his contributions to the Institute
and the University. All three events shared a friendly atmosphere
and productive discussions between the participants.

The LQG workshop entitled ``Frontiers of Loop Quantum Gravity" had
25 participants and the discussions centered among five topics of
current interest: Phenomenology, Semiclassical Issues, Hamiltonian
Constraint, Spin Foams and Future Directions. Each session was
three hour long, with 2 short presentations and lots of
discussion. On Friday Jan 30th, the workshop started with a
presentation by Daniel Sudarsky on recent results that show that a
low energy effective description in terms of a Lorentz violating
theory (with a preferred frame) is already ruled out by existing
experimental and theoretical bounds. This leads to a subtler
possibility for manifestations of the discreetness coming from the
Planck scale in the form of a Double Special Relativity (DSR)
framework. L. Freidel gave a review talk on the two kinds of DSR
theories which he called DSR-1 and DSR-2, and their prospects for
becoming physically viable.

The Session on Semiclassical issues had two short presentations.
Hanno Sahlmann gave a nice summary of the current approaches to
construct semi-classical states of the theory, from the now
classical {\it weave} states,  gauge coherent states and shadow
states to statistically generated states. Abhay Ashtekar gave a
presentation on possible physical applications that the
semiclassical states might try to attack; in particular the
construction of states that approximate the Minkowski vacuum and
the definition of effective potentials for scattering scenarios.

Spin Foams was the topic of the next session on Sat. Jan 31st,
where Alejandro Perez gave a status report on the Barret Crane
model. In particular, there was discussion on recent progress on
the proper definition of the path integral measure for the
Plebanski action and its possible relation to the elimination of
bubble divergences when the symmetries of the system are properly
addressed. The other important issue that was discussed was the
dominance of degenerate configurations in the asymptotics of the
terms that compose the state sum model. Louis Crane discussed
possible new mathematical descriptions for spin foam models via an
n-categorical approach.

In the Hamiltonian Constraint session, Abhay Ashtekar discussed
the  status of the so-called Thiemann-like Hamiltonians. Are they
clinically dead? The consensus after the discussion was that they
are still alive, and no fatal flaw has been found in the approach,
even when there is still no agreement on the right way of finding
a projector on physical states. Thomas Thiemann gave an
introduction to the Master Constraint program, where the
particular proposal for finding such operator seems to be very
promising. It was agreed upon that this approach might shed light
on defining the projector (via a spin foam model).

In the Morning of Sun Feb 1st Jorge Pullin gave a summary of the
results that he and R. Gambini have found in defining consistent
discretizations for  quantum and classical gravity. In particular,
he discussed some of its applications to quantum cosmology and
decoherence. Abhay Ashtekar summarized what had been discussed in
the workshop in the Concluding Remarks, and a general discussion
followed. In particular, there was discussion about different
possible discrete frameworks, observables associated to particles
and the cosmological constant in $2+1$ gravity and the relation
between spin foams and the continuum.

The general agreement between the participants was that the
meeting was very productive for clarifying (and reaching
consensus) on several conceptual and technical issues, and in
defining direction in which we should focus our attention in the
future. There was such excitement that this was dubbed the ``First
NAFTA meeting on LQG" and the participants somewhat committed to
followups to this workshop that would take place in Canada, the
USA and M\'exico on a rotating basis.

On Monday Feb 2nd, the Ryan-fest was held at UNAM, to celebrate
both Mike's 60th birthday and 30 years of his arrival to M\'exico.
The program was divided into technical talks and more informal
`anecdotic' ones: Abhay Ashtekar gave a seminar on Loop Quantum
Cosmology followed by a talk on Non-commutative Quantum Cosmology
by Octavio Obregon and later on by Richard Matzner on
Linearizations and Hyperbolicity in Numerical Relativity. On the
less technical side there were talks by Marcos Rosenbaum who
described Mike's path during this last 30 years and his influence
in starting and consolidating the gravity group at UNAM. Roberto
Sussman gave us a student's perspective of Mike as a teacher in
the `old times', and Jorge Pullin took us to a Journey of
discovery through the eyes of Mike's papers of all times.

The final event was a two-day celebration for Marcos Rosenbaum who
was responsible for the mere existence of the Institute for
Nuclear Sciences (and the Gravity Department) as a research
center, of which he was the director for 16 years. There were two
kinds of talks: those by present and past University Officers
(including the present and a former Rector), who highlighted Prof.
Rosenbaum's service to the University at large. The second series
of talks were of a scientific nature and were divided into the two
main areas of research of Prof. Rosenbaum: General Relativity and
Mathematical Physics. Among the invited speakers were Stanley
Deser, Richard Matzner, Octavio Obregon and Raymundo Bautista.
These talks were complemented by talks from `local' people such as
M. Alcubierre, C. Chyssomalakos, M. Ryan, A. Turbiner and the
author of this note.

What this series of events made clear for the participants was
that the Gravity Department at UNAM
\htmladdnormallink{http://www.nuclecu.unam.mx/\~{}gravit}
{http://www.nuclecu.unam.mx/\~gravit}, after only 30 years of its
birth, is now coming of age, fit and running. We wish the
department, and both Professors Ryan and Rosenbaum many more years
of healthy and productive life.

\vfill\eject
\section*{\centerline {
Building bridges: CGWA inaugural meeting in Brownsville}}
\addcontentsline{toc}{subsubsection}
{\it Building bridges, by Mario D\'{\i}az}
\begin{center}
Mario Diaz, UT Brownsville
\htmladdnormallink{mdiaz@utb1.utb.edu}
{mailto:mdiaz@utb1.utb.edu}
\end{center}

Building Bridges, as it was called the inaugural meeting of the Center
for Gravitational Wave Astronomy at the University of Texas at
Brownsville was held in Brownsville at the school main campus on
December 14 and December 15 of last year.

The title was a suggestive metaphor to indicate the main goal of the
CGWA, which is to give a common ground to the efforts of data
analysts, source modelers and astrophysicists in support of
gravitational wave astronomy and in particular of LISA science. The
CGWA was created with NASA support under the agency's university
research centers program.  The program of the conference consisted of
several invited talks and a contributed poster session.

The first day Bernard Schutz (AEI) and Tom Prince (Caltech-JPL) talked
about the science promise of LISA and the LISA challenges for the
scientific community, respectively. Schutz described in detail several
aspects of the LISA mission, its challenges and its promises. He
pointed out how different other presentations through the meeting were
resonating with the major themes for LISA, i.e.: Supper Massive Black
Holes (SMBH), how have they accompanied galaxy formation and their
merger history, LISA noise confusion problems and the data analysis
needs for the mission. He also described the future beyond LISA and
other missions that are in the design stages like the Big Bang
Observer. Prince framed LISA's mission in the context of NASA's
structure and evolution of the universe roadmap. He referred to the
significance within this context of the agency's Beyond Einstein
program, which concentrates on understanding what powered the big
bang, what happens at the edge of a black hole and what is dark
energy.

Then Doug Richstone (U. of Michigan) completed the morning session
with his talk about ``Things invisible to see: Supermassive black holes
in ordinary galaxies". Richstone concentrated on discussing what has
triggered the interest on them, the current demographic picture and
emerging developments in the field. He particularly discussed the
possibility of gravitational wave observations of black hole
mergers. 

In the afternoon David Merritt (Rutgers) and Simon Portegeis Zwart
(U. of Amsterdam) discussed the astrophysics of black holes, while
Neil Cornish (Montana State) and Lior Barack (CGWA-UTB) referred to
different aspects of LISA science.  Merritt talked about the
Astrophysics of Binary Supermassive Black holes. His presentation
highlighted some of the outstanding questions in the field, like the
resolution of the final parsec problem, and other different aspects of
the dynamical evolution of black holes (are there un-coalesced binary
black holes? how are black holes ejected from galaxies?  Zwart
discussed in detail the technique of n-body simulations in regards to
the formation of intermediate mass black holes in young dense star
clusters. Cornish discussed with some extension LISA challenges in the
area of data analysis. He presented some approaches to the solution of
some of these challenges with the utilization of an array of
techniques that have been developed by his group. Barack started with
a quick overview of capture sources and data analysis problems
associated. He then introduced a class of approximate analytic
waveforms and showed how can be utilized to estimate LISA's parameter
extraction accuracy and to estimate SNR thresholds for detection. He
revisited detection rates in light of this model as well.  

A banquet followed in the evening with the normal occasion for
socialization. Bernard Schutz talked about the significance of the
creation of the CGWA and Richard Price referred in a jovial tone to
the accomplishments and future endeavors of gravitational wave
physicists in Brownsville.

In the morning of the following day Beverly Berger (NSF) started with
a presentation about NSF role in support of gravitational wave science
in particular and gravitation and general relativity at large.  Soumya
Mohanty (CGWA-UTB) presented ``Beyond the Gaussian, stationary
assumption: data analysis techniques for real interferometric data".
The following talks concentrated on issues and challenges in the area
of numerical source simulation: Carlos Lousto (CGWA-UTB) spoke on
``What can be learned about binary clack hole evolution due to
gravitational radiation". Bernd Bruegmann (PSU) discussed several
issues related to black hole simulations in numerical relativity.
Richard Price (U. of Utah) spoke on the ``Periodic Standing Wave"
method, an alternative to full numerical relativity for binary black
hole inspiral. This method uses an exact numerical solution for
rigidly rotating (``helically symmetric'') sources and fields to give an
approximation for the intermediate epoch of inspiral in which the
holes have strong gravitational interaction, but are not yet in their
final plunge.  Luis Lehner (LSU) made an effort at predicting the next
steps in the simulations of Einstein's equations, while John Baker
(GSFC) talked about the Lazarus approach.  Yasushi Mino (CGWA-UTB)
gave the last lecture of the conference on the past, present and
future of the self-force problem.  The slides for the talks can be
viewed at the CGWA web site:
\htmladdnormallink{http://cgwa.phys.utb.edu/events/program.php}
{http://cgwa.phys.utb.edu/events/program.php}.

The contributed poster session was also well attended and consisted in
a good number of posters covering a wide range of topics within the
meetings theme.  A very positive aspect to remark from this meeting is
the large number of students that attended from different places in
the country and also from abroad.

\vfill\eject
\section*{\centerline {Strings Meet Loops}}
\addcontentsline{toc}{subsubsection}{\it 
Strings Meet Loops, by Martin Bojowald}
\begin{center}
Martin Bojowald, Albert-Einstein-Institute, Potsdam, Germany
\htmladdnormallink{mabo@aei.mpg.de}
{mailto:mabo@aei.mpg.de}
\end{center}

During October 29--31, 2003 the Albert-Einstein-Institute in Potsdam,
Germany hosted the Symposium ``Strings Meet Loops'' which was
organized by Abhay Ashtekar and Hermann Nicolai.  The primary purpose
was to bring together researchers working on string theory on the one
hand, and on canonical and loop quantum gravity on the other, and to
enhance the exchange of ideas between the two
communities. Correspondingly, the program consisted of talks that were
primarily addressed to the other community: four from string theorists
(Kasper Peeters, Bernard de Wit, Michael Douglas and Jan Plefka) and
four from the loop community (Abhay Ashtekar, Jurek Lewandowski,
Martin Bojowald and Laurent Freidel). They provided broad overviews of
the current areas of active research in both approaches, focusing on
conceptual frameworks and physical issues.  In addition, there were
two talks of interest to both programs but not explicitly belonging in
either (Marc Henneaux and Max Niedermaier) as well as Introductory
Remarks by Hermann Nicolai and Closing Remarks by Abhay Ashtekar. Over
50 participants from Europe, US and Canada attended the symposium. In
addition, researchers from the Perimeter Institute and Rutgers
participated in the afternoon sessions via video camera. (The full
program and images of the transparencies used in the talks will remain
available for download from the web page
\htmladdnormallink{http://www.aei-potsdam.mpg.de/events/stringloop.html}
{http://www.aei-potsdam.mpg.de/events/stringloop.html}.)

The main purpose of all the talks was to serve as concrete
platforms for subsequent discussions for which plenty of time was
allotted by the organizers and used by the participants. In fact,
the discussions quickly extended from the material covered in the
preceding talk to its general area, resulting in a lively exchange
of viewpoints from the different perspectives represented by
members in the audience. In this regard, the hopes of the
organizers were exceeded by the vibrant atmosphere during the
symposium which resulted in frank discussions of the main open
problems and of expectations toward the other community concerning
issues which should be addressed in the future. Several prevalent
misunderstandings have been clarified in these discussions.

The success of the symposium can also be seen in the interest from
different sides to organize a a second symposium ``Strings Meet
Loops: 2,'' in the future.

\vfill\eject

\section*{\centerline {
International Conference on Gravitation and Cosmology 2004}}
\addcontentsline{toc}{subsubsection}{\it  
ICGC -- 2004, by Ghanashyam Date}
\begin{center}
Ghanashyam Date, Institute of Mathematical Sciences, Chennai, INDIA
\htmladdnormallink{shyam@imsc.res.in}
{mailto:shyam@imsc.res.in}
\end{center}

The {\em International Conference on Gravitation and Cosmology} is a
series of conferences held approximately every four years in India. The
series has been conceived as a means to provide younger researchers an 
exposure to latest research trends and to  promote interaction between the 
International and the Indian research communities. Each of these conferences
focuses on two or three `theme topics' and are typically attended by about 
a 100 participants from India and abroad. The previous conferences 
in this series were held at Goa (1987), Ahmedabad (1991), Pune (1995) and
Kharagpur (2000) and have been quite successful in the stated
objectives. 

The fifth conference in the ICGC series, ICGC-2004, was
organized by the Cochin University of Science and Technology
(CUSAT)   at the {\em
Riviera Suites} on the outskirts of Cochin
during January 5--10, 2004. It had 17 plenary talks and, as a new feature,
it also had 8 short talks which were more specialized than the plenary
talks but still accessible to a wider audience. There were three focus
themes: Cosmology, Gravitational Waves and Quantum Gravity.
About 70 contributed papers were presented in oral presentations and
poster sessions in four workshops on: Quantum
Aspects of Gravitation, Classical Aspects of Gravitation, Cosmology
and Gravitational waves and Relativistic astrophysics. 

{\underline {Cosmology:}} {\em Robert Crittenden} summarized the WMAP results
and {\em Manoj Kaplighat} discussed the early re-ionization aspect. {\em 
Jerry Ostriker} gave a status summary of the `standard model' of cosmology 
and emphasized the need and role of various complementary observations in 
building up a comprehensive picture of cosmology. Looking some what into 
the future, {\em Subhabrata Majumdar} discussed cluster surveys while
{\em Bhuvnesh Jain} discussed weak lensing. Finally, {\em Robert
Crittenden} doubled up for {\em Edmund Copeland} and surveyed a variety 
of theoretical ideas being pursued, some quite desperate, regarding an
understanding of the Dark Energy. 

{\underline {Gravitational Waves:} Talks in this theme covered various
aspects from analytical and numerical computations of wave forms to
interesting design aspects of interferometric detectors. 
 {\em Gabriela Gonz\'alez} described the LIGO
experiment and the science runs. {\em Vicky Kalogera} discussed the
event rate estimations from compact binaries. Detector assembly
integration and simulations were discussed by {\em Biplab Bhawal} while
{\em Sanjeev Dhurandhar} discussed in detail data
analysis strategies required to construct efficient and effectual
templates for gravitational wave detection. 
The analytical computations of the chirps within the PN
expansion framework was discussed by {\em Luc Blanchet}. {\em Masaru
Shibata} provided the current status and an optimistic future of numerical 
simulations using supercomputers. {\em Frederic Rasio} discussed the 
possibility of constraining the equation of state for neutron stars from 
observations of gravitational wave forms. 

{\underline {Quantum Gravity:}} In this theme there was one talk on string
QG, two on loop QG, two on black hole entropy in various approaches,
one on brane cosmology and one on phenomenological QG. {\em Sandip
Trivedi} described the recent advances in getting not one but a very
large number ($\sim 10^{100}$) of De Sitter vacua in string theory. This 
is achieved by a controlled SUSY breaking in a compactification with fluxes.
{\em Jorge Pullin} briefly described the main historical steps in the
loop quantum gravity program and then focused on the recent proposal of
a priori discretization of space-time. 
In a discrete time formulation,
the evolution is achieved by finite canonical transformations. For a
constraint theory, this allows one to obtain a constraint-free
formulation. Various implications were discussed. 
{\em Martin Bojowald}
detailed the loop quantum cosmology framework and presented interesting recent
results. {\em Saurya Das} presented a fairly comprehensive comparison of
the calculation of black hole entropy in various approaches such as
strings, LQG, horizon CFT, AdS/CFT etc. including the logarithmic
corrections. He also discussed the attempts to understand the  Hawking effect
and the  information loss issue. {\em Parthasarathi Majumdar} discussed the
universality of canonical entropy including logarithmic corrections.
Brane cosmology, particularly the possibility of inflation being
driven by a scalar field propagating in the bulk as well, was discussed
by {\em Misao Sasaki}. {\em Jorge Pullin} summarized the attempts to look 
for QG signals via Lorentz invariance-violating modifications of the 
dispersion relations. 

{\underline {Other talks:}} Apart from the talks devoted to the main
themes, there were five talks dealing with different aspects of
gravitation. {\em Clifford Will} summarized the current limits on the
various parameters from the traditional tests of GR and also discussed
newer possibilities from gravitational waves. Gravitational collapse and
the status of naked singularities was summarized by {\em Tomohiro
Harada}. {\em Patrick Das Gupta} discussed the so called `short'
duration GRB's and their statistics. {\em Sayan Kar} discussed the issue
of quantification of `small' violations of the averaged energy condition
and its role in traversable worm holes. The classic topic of `Kerr-Schild
geometries' was discussed by {\em Roy Kerr}. {\em Ghanashyam Date}
gave an over view of the conference.

There was also an `outreach lecture' by {\em Clifford Will} on {\em
``Was Einstein Right?"} for the undergraduate students in a city
college. The otherwise intense atmosphere of the conference was
lightened by a delightful pre-dinner talk by {\em C. V. Vishveshwara}
titled: ``Cosmos in Cartoons".

The conference had a half-a-day `cultural session' consisting of a {\em
backwater cruise}, the traditional classical dance of {\em Kathakali}
(narration of stories through dance) and the, unique to Kerala,
twelfth-century martial arts called {\em Kalaripayat} followed by the 
conference banquet.

After this memorable experience, one looks forward to future
ICGC meetings.

Links to presentation as pdf/ppt/ps files can be soon found at 
Presentation (pdf/ppt/ps files) will soon be available at 
\htmladdnormallink{http://meghnad.iucaa.ernet.in/\~{}iagrg/talks.html}
{http://meghnad.iucaa.ernet.in/\~iagrg/talks.html}.

The proceedings of the conference will be published as a special issue
of {\em Pramana -- Journal of Physics}.

The conference was sponsored by: The Abdus Salam ICTP, Italy; BRNS
(DAE), Mumbai; CSIR, New Delhi; DST, New Delhi; HRI, Allahabad; IIA,
Bangalore; ISRO, Bangalore; IMSc, Chennai; IOP, UK; IUCAA, Pune; RRI,
Bangalore and UGC, New Delhi.

\vfill\eject
\section*{\centerline {
27th Spanish Relativity Meeting (ERE-2003)}}
\addcontentsline{toc}{subsubsection}{\it  
27th Spanish Relativity Meeting (ERE-2003) by Juan Miralles, Jose Font, Jose Pons} 
\begin{center}
Juan A. Miralles (Alicante), Jos\'e A. Font (Valencia), Jos\'e A. Pons
(Alicante)
\htmladdnormallink{ja.miralles@ua.es}
{mailto:ja.miralles@ua.es},
\htmladdnormallink{ja.miralles@ua.es}
{mailto:j.antonio.font@uv.es},
\htmladdnormallink{jose.pons@ua.es}
{mailto:jose.pons@ua.es}
\end{center}

The 27th Spanish Relativity Meeting (ERE-2003) was held at the
University of Alicante (Spain) from September 11 to 13, 2003. The
Spanish Relativity Meetings are annual conferences on General
Relativity which started back in 1977, each year being organized by
one of the different groups doing research on Relativity and
Gravitation in Spain.  In this occasion the meeting was jointly
organized by groups from the University of Alicante and the University
of Valencia. A brief history of these meetings can be found online at
the conference web site 
\htmladdnormallink{http://www.sri.ua.es/congresos/ere2003/ereeng.htm}
{http://www.sri.ua.es/congresos/ere2003/ereeng.htm}

The scope of the conference has expanded over time as a follow up of
research activity in Relativity and Gravitation in Spain. The program
of the 27th edition included a series of six plenary lectures in the
mornings, plus several shorter communications, both in the morning and
in the afternoon, where participation of young scientists was, as
usual, encouraged. The number of delegates attending the meeting was
over sixty people, mostly from Spain but also from abroad. A total
number of thirty-three contributed talks were presented, covering a
broad range of topics including cosmology, numerical relativity, and
formal aspects of mathematical relativity. The subjects of the plenary
lectures covered topics of current interest in the field of
Gravitational Radiation. The abstracts of these talks can be accessed
online at the conference web site
\htmladdnormallink{http://www.sri.ua.es/congresos/ere2003}
{http://www.sri.ua.es/congresos/ere2003}.

Ewald M\"uller (MPA, Germany) gave a review talk on core collapse
supernovae as sources of gravitational radiation. In particular he
showed pioneer results on the gravitational wave emission from
simulations of highly aspherical models on which the asphericities are
caused by convective mass flow both in the proto-neutron star and in
the post-shock neutrino heated hot bubble region.

In his talk Nils Andersson discussed the various ways in which neutron
stars may give rise to detectable gravitational waves. He described
the modeling of these systems which requires an understanding of much
of modern physics, ranging from general relativity to superfluidity
and nuclear physics at extreme densities. The main aim of his talk was
to outline how gravitational-wave data can be used as a probe of
exotic physics in neutron stars.

Signal detection was discussed in the corresponding talks of Alberto
Lobo (University of Barcelona), Alicia Sintes (University of the
Balearic Islands and AEI), Pia Astone (University of Rome), and
B. Sathyaprakash (Cardiff University). In particular Pia Astone
discussed recent controversial results obtained with two gravitational
wave resonant bar detectors, Explorer (located at CERN) and Nautilus
(in Frascati, LNF).  These detectors are allowing to investigate
various classes of signals, such as bursts, continuous waves,
stochastic background. They operated in the year 2001 with
unprecedented sensitivities, being potentially able to detect the
conversion of $10^{-4}$ solar masses in the Galaxy into gravitational
waves.

Alicia Sintes presented a talk focused on the search of continuous
gravitational waves from rotating neutron stars, which are among the
most promising sources for ground based interferometer
detectors. Although young rapidly rotating neutron stars are probably
better initial candidates for gravitational wave detection than the
known set of radio pulsars, the data analysis problem for these
putative sources is more difficult because of their unknown location
and frequency evolution. Since the expected gravitational wave
amplitude from pulsars is very weak, it is necessary to integrate the
data for long periods of time (months to year) with the
signal-to-noise ratio increasing roughly as the square root of the
observing time. Several data analysis techniques have been used, and
others are under development, which are able to handle efficiently
these long stretches of data. These techniques were thorough fully
discussed in her talk.

The problem of searching for black hole binaries was discussed in the
talk by B.  Sathyaprakash. Post-Newtonian calculations and numerical
relativity simulations have provided with waveform templates that can
be effectively used to identify the inspiral and merger signals buried
in noisy data. Sathyaprakash presented the state-of-the-art techniques
of search algorithms tailored to dig out signals of known shape, which
have been developed and implemented to increase the chance of
detection. Applying these waveform templates and search algorithms on
real data have taught us important lessons about how to deal with
problems arising from handling non-stationary and non-Gaussian
backgrounds.

Alberto Lobo gave a review talk on LISA, the first space borne
gravitational wave detector, a joint ESA-NASA mission scheduled for
launch in 2011. He presented the scientific objectives of LISA, as
well as its present development status. Lobo also described the
situation of the Spanish involvement in the mission.

Finally, part of the meeting was devoted to discussing the
establishment of the Spanish Society of Relativity and Gravitation
(SEGRE). The 28th edition of the Spanish Relativity Meeting will take
place in Madrid from September 22 to 24, 2004.

\vfill\eject

\section*{\centerline {
Mathematical Relativity: New Ideas and Developments
}}
\addcontentsline{toc}{subsubsection}{\it  
Mathematical Relativity: New Ideas and Developments, by Simonetta Frittelli}
\begin{center}
Simonetta Frittelli, Duquesne University,
\htmladdnormallink{simo@mayu.physics.duq.edu}
{mailto:simo@mayu.physics.duq.edu}
\end{center}
\parskip=3pt

This meeting took place in Bad Honnef, on the shores of the Rhine, Germany, on
March 1-5, 2004. This was the latest in an ongoing series of seminars sponsored
by the Heraeus foundation, the most important German private funding
institution in physics and one that cooperates frequently with the German
Physical Society.

The meeting was principally aimed at a broad audience that included a large
representation of junior participants. There were 60-minute plenary talks in
the morning, and 30-minute selected contributed talks in the afternoon. The
plenary talks covered mostly recent developments in topological, analytical and
numerical methods in mathematical relativity. Undoubtedly by thoughtful design
on the part of the organizers (J\"{o}rg Frauendiener, Domenico Giulini and Volker
Perlick), many plenary talks complemented or supplemented each other in a
refreshingly cohesive way.

The conference was opened by Bobby Beig (Vienna) with a description of
hyperbolic equations for matter fields.  This was followed by a presentation on
the space of null geodesics by Robert Low (UK), and a discussion about the
geometry of pp-wave spacetimes (which generalize classical plane waves) by
Miguel Sanchez (Spain).  

In what was characterized as ``the revenge of the physicists'' by some
participants, the presentations of the second day all revolved around the
interplay between numerical and analytical issues, with talks delivered by Jeff
Winicour, myself, Luis Lehner and Beverly Berger (all USA). These talks were
meant to illustrate not only how analysis and geometry can guide numerical
efforts, but, conversely, how numerical simulations can provide clues for
analytic developments.  

Bob Wald (USA) opened the third day with a review of conserved quantities
associated with hypersurfaces that go to null infinity in asymptotically flat
spacetimes, which prompted a curious dialogue between Bob and Roger Penrose
(UK) about what ``normal'' people refer to as the ``problem'' of angular momentum.
This was followed by the revenge of the mathematicians Paul Ehrlich (USA), with
an anecdotic perspective on the influence of the book {\it Global Lorentzian
Geometry}, and Antonio Masiello (Italy), who spoke about the Avez-Seifert
theorem for the relativistic Lorentz equation.  The day ended with an evening
talk by Roger Penrose about the three {\bf F}'s in modern physical theories:
{\bf F}ashion (strings), {\bf F}aith (quantum mechanics) and {\bf F}antasy
(inflation).    

Undaunted by the implications of Penrose's address, Alan Rendall (Germany)
delivered an impression of good old inflation from a mathematician's point of
view the very next day, followed by Helmut Friedrich (Germany) on conformal
infinity and its connection to the Newman-Penrose constants, and Sergio Dain
(Germany) with a description of the use of trapped surfaces as boundaries for
the constraint equations on initial values for black hole spacetimes. 

The meeting closed with talks by Greg Galloway (USA) on asymptotically deSitter
spacetimes, and Laszlo Szabados (Hungary) on the definition of center of mass
associated with Cauchy surfaces in asymptotically flat spacetimes. 

The venue for the meeting was highly conducent to interaction between
junior and senior participants. This was a large mansion (the Physikzentrum)
featuring accomodations, dining and a state-of-the-art conference room. There
was a common room where breakfast and lunch were served at tables for six
people, and an old-fashioned basement cellar where a buffet dinner was
available and where much informal social interaction took place in the evenings
(apparently including fortune readings).  A conference picture can be accessed
at
\htmladdnormallink{http://mayu.physics.duq.edu/\~{}simo/badhonnef04.html} 
{http://mayu.physics.duq.edu/~simo/badhonnef04.html}

The substance of the the plenary talks will be published as a Springer volume
in the near future. The complete program including the abstracts of the
contributed talks and posters is available online at

\htmladdnormallink{http://www.tat.physik.uni-tuebingen.de/\~{}heraeus/}
{http://www.tat.physik.uni-tuebingen.de/~heraeus/}.
\parskip=10pt

\end{document}